\documentclass[prl,twocolumn]{revtex4}

\newcommand{\calH}{{\cal H}}
\newcommand{\calA}{{\cal A}}

\usepackage{graphicx}
\usepackage{bm}

\begin{document}

\title{Direct Perturbation Theory on the Electron Spin Resonance Shift and Its Applications}

\author{Yoshitaka Maeda and Masaki Oshikawa}
\affiliation{
Department of Physics, Tokyo Institute of Technology, \\
Oh-oka-yama, Meguro-ku, Tokyo 152-8551, Japan
}

\begin{abstract}
We formulate a direct and systematic perturbation theory on the shift
of the main paramagnetic peak in electron spin resonance,
and derive a general expression up to the second order.
It is applied to the one-dimensional XXZ and transverse Ising models
in the high field limit,
to obtain explicit results including the polarization dependence
at arbitrary temperature.
\end{abstract}



\maketitle

\noindent {\em Introduction --- }
Electron spin resonance (ESR)
is not only an important experimental method,
but also poses fundamental problems in theoretical physics.
At the same time, it provides a good testing ground for theories,
as very precise spectra can be obtained in many experiments.
This is evident historically in the
fact that the general formulation of linear response theory,
the so-called Kubo formula, appeared first in a treatment
of magnetic resonance by Kubo and Tomita,~\cite{KT} and this became
a standard theory on ESR.
In addition to steady advances in experimental techniques,
a recent development of a field theory formulation~\cite{OshikawaAffleck,OshikawaAffleck-PRB}
for one-dimensional systems
and also of numerical approaches~\cite{Miyashita,Iitaka,Maeda:XY}
renewed interest in ESR.

In ESR, we study a magnetic system under an applied field $H$.
The direction of the applied field is defined as $z$ axis. The Hamiltonian
is given by
\begin{equation} 
 \calH = \calH_Z + \calH_{SU(2)} + \calH',
\label{eq:fullHam}
\end{equation}
where $\calH_{SU(2)}$ represents the isotropic interactions such as the Heisenberg
exchange, and $\calH'$ is the anisotropy which is often small.
$\calH_Z = - H S^z$ is the Zeeman term, and the notation $S^\alpha$
is used for the total spin operator
as $S^{\alpha} \equiv \sum_j S^{\alpha}_j$
throughout this Letter.

An electromagnetic radiation, with the frequency $\omega$ of the same order as
$H$, is applied and its absorption is measured.
We consider the Faraday configuration, in which the oscillating
magnetic field is perpendicular to the $z$ axis.
Generally, within the linear response theory, the absorption intensity
is related to the imaginary part $\chi''_{aa}(\omega)$
of the complex dynamical spin susceptibility
({\it i.e.} the retarded Green's function of the spin operators)
$\chi_{aa}(\omega)$
at zero momentum,
where $a$ is a spin axis determined by the polarization
of the electromagnetic wave.
If the oscillating magnetic field is parallel to the $x$ ($y$) direction,
the absorption intensity is related to $\chi''_{xx}$ ($\chi''_{yy}$).
In the presence of general anisotropy, $S^x$ and $S^y$ are not
equivalent, since they give different spectra. We call this dependence polarization dependence~\cite{Natsume,OshikawaAffleck,OshikawaAffleck-PRB}.

In this Letter, for simplicity,
we discuss $\chi''_{aa}(\omega)$ as a function of $\omega$
at a fixed $H$, although usually in experiments
the absorption intensity $\propto \omega \chi''_{aa}(\omega)$
is measured with varying $H$ while $\omega$ is constant.
When the spins are non-interacting, $\calH = \calH_Z$,
we trivially obtain
\begin{equation}
\chi''_{aa}(\omega) \propto \delta(\omega - H).
\label{eq:delta}
\end{equation}
We thus find the delta-function peak, which is called the paramagnetic
resonance, at $\omega=H$.
A characteristic of ESR, which measures the zero wave vector component
of $\chi''$, is that the lineshape still keeps the same form
even in the presence of $\calH_{SU(2)}$.
Nontrivial changes in the lineshape only occurs in the presence
of anisotropy.
For a small anisotropy $\calH'$,
the paramagnetic resonance at $\omega=H$ is often shifted and broadened.
It is thus useful to obtain the shift and width of the main paramagnetic
resonance peak in terms of the perturbation theory in the anisotropy
$\calH'$, with the unperturbed Hamiltonian
$\calH_0 = \calH_Z + \calH_{SU(2)}$.

However, this is not as easy as it may sound, even in a weakly interacting
system. A realistic, broadened lineshape cannot be obtained
in any finite order of the perturbative expansion of $\chi''$.
This is apparent as $\chi''$ has
the delta function at the zeroth order, while the ``true'' $\chi''$ with
a broadened resonance generally does not contain a delta function. The delta
function could only be eliminated by a summation of the perturbation
expansion of $\chi''$ up to the infinite order.

Several existing approaches to the shift and width are based on
an infinite summation of the perturbation series, relying on various
grounds.
In the Kubo-Tomita theory~\cite{KT}, the infinite summation
is carried out by assuming that higher order terms in the cumulant expansion vanish.
Although this assumption appears to be correct in certain cases,
its range of validity has not been clarified.
Indeed, it was pointed out recently that a straightforward application
of the Kubo-Tomita theory fails in the presence of the
Dzyaloshinskii-Moriya interaction~\cite{Choukroun,OshikawaAffleck,OshikawaAffleck-PRB}.
For $S=1/2$ chain at low temperature, a field theory approach based on
bosonization technique is developed~\cite{OshikawaAffleck,OshikawaAffleck-PRB}.
There, the infinite summation is performed systematically in terms of the Feynman-Dyson self-energy. 
However, this is only possible because of the particular
mapping of the ESR problem to the Green's function of the boson field,
which holds only
for one-dimensional system at low temperatures.
Therefore, no immediate generalization of
this technique to other cases of interest is possible.

In the Mori-Kawasaki theory~\cite{MK},
an ansatz is made for the dynamical spin susceptibility as
\begin{equation}
\chi_{aa}(\omega) \propto \frac{-1} {\omega - H - M(\omega)},
\label{eq:MKansatz}
\end{equation}
in terms of the memory function $M(\omega)$, which is
assumed to be smooth at $\omega \sim H$.
It is actually of the same form that results from the
self-energy summation.
With this ansatz,
it is possible to determine the shift
and width from the low order perturbative expansion of the
susceptibility.
However, it should be noted that in the absence of the diagrammatic
perturbation theory,
there is generally no justification of
the memory function ansatz~(\ref{eq:MKansatz}).
Thus, its range of validity has not been clarified.

The shift and width, on the other hand, have also been discussed
in terms of the moments.
The ESR shift may be related to the normalized first moment that is
the ``mean frequency'' of the entire ESR absorption.
However, it often happens that, in the presence of the anisotropy,
the peaks of the absorption spectrum appear at higher frequencies such as
around $2H, 3H, 4H \ldots$.
In the presence of such satellite peaks,
in order to discuss the shift of the main paramagnetic resonance to
higher orders,
one must calculate the projected moments which are defined by restricting the
domain of integration to the support of the main paramagnetic resonance.
However, the existing
calculations~\cite{VanVleck,Pryce,Usui,KanamoriTachiki,McMillan,McMillan_J_60,McMillan_J_61}
of the projected moments
involve nontrivial assumptions, and their range
of validity is again questionable.
The polarization dependence has not been discussed
in terms of these approaches based on the projected moments.

Given this situation, 50 years after the publication of the Kubo-Tomita theory,
it would be worthwhile to develop an alternative approach to ESR
which is complementary to the existing ones.

\noindent {\em Formulation ---}
In this Letter, we aim to develop a perturbation theory directly on
the shift of the main paramagnetic resonance.

For simplicity, we only consider a linear polarization of
the oscillating magnetic field.
When the oscillating magnetic field is parallel to the $a$ axis,
the ESR peak position $\omega_{m,a}$  
may be defined by a maximum of $\chi''_{aa}$, which satisfies
\begin{equation}
 \partial_{\omega} \chi''_{aa}(\omega_{m,a}, \alpha) = 0,
\label{eq:max}
\end{equation}
where $\alpha$ is the small perturbation parameter which gives the
magnitude of the anisotropy $\calH'$.
For the sake of brevity, we will omit the polarization index $a$
from $\omega_{m,a}$ until we discuss the polarization dependence.

Our strategy is to expand the peak frequency in $\alpha$ as
\begin{equation}
 \omega_m = \omega_m^{(0)} + \alpha \omega_m^{(1)}
  +\alpha^2 \omega_m^{(2)} + \ldots .
\label{eq:expomega_m}
\end{equation}

However, an attempt to solve condition~(\ref{eq:max})  perturbatively
encounters a problem already at the zeroth order.
This is because, at the zeroth order,
the dynamical susceptibility is given by the singular delta function
as in~(\ref{eq:delta}). In fact, at the zeroth order, (\ref{eq:max})
holds for any $\omega_m \neq H$.

The delta function can be defined as
\begin{equation}
 \delta(\omega -H)  = -\frac{1}{\pi}
\lim_{\epsilon \rightarrow +0}{\rm Im} \frac{1}{\omega - H +i \epsilon},
\end{equation}
where $\epsilon$ is often called the convergence factor, which is
usually regarded as positive infinitesimal.
The introduction of the convergence factor corresponds to
multiplying $e^{-\epsilon t}$ with the retarded Green's functions,
as functions of real time $t$.
Generally, the inclusion of the convergence factor amounts
to replacing $\omega$ with $\omega + i \epsilon$.

In order to circumvent the above problem due to the singularity,
we perform the perturbative expansion~(\ref{eq:expomega_m}),
keeping the convergence factor $\epsilon$ finite.
The limit $\epsilon \rightarrow +0$ is taken after the expansion
is obtained.
In fact, a finite $\epsilon$ is often used
to estimate the continuous spectral functions from a finite-size
numerical calculation,
which inevitably yields a collection of delta functions.
Sometimes, $\epsilon$ may be considered as a small but finite
parameter that represents the relaxation effects which do not appear in
the microscopic model but actually exist in a real system.

If the shift is Taylor expandable in both $\epsilon$ and $\alpha$,
this approach would give the correct perturbation series
in $\alpha$ after taking the $\epsilon \rightarrow 0$ limit.
This should be the case when 
the Mori-Kawasaki memory function ansatz is valid, as
the shift is given by the real part of the memory function,
which is assumed to be smooth at $\omega \sim H$ and to be
Taylor expandable in $\alpha$. As we will see below, our formulation
is actually more general than the Mori-Kawasaki ansatz.
Namely, our approach gives correct answer even when the line shape is
Gaussian. In contrast, the ansatz (\ref{eq:MKansatz}) does not match the
Gaussian line shape.

On the other hand, when the main paramagnetic resonance splits
into multiple peaks, there would be multiple solutions
to eq.~(\ref{eq:max}) at $\omega \sim H$.
Thus the present argument
(as well as the Mori-Kawasaki memory function ansatz) does not
apply to such cases in a straightforward manner.
The generalization to this class of problems is discussed
elsewhere~\cite{inprogress}.
In this Letter, we focus on the simplest case
with the single main paramagnetic resonance peak.

Equation (\ref{eq:max}) is then expanded as
\begin{eqnarray}
\label{eq:expansion}
0&=&\partial_{\omega} \chi^{''}_{aa}(\omega_m,\alpha)\nonumber\\
&=&\partial_{\omega} \chi^{''}_{aa}(\omega_m^{(0)},0) \nonumber \\
&&
+\alpha\{\partial^2_{\omega} \chi^{''}_{aa}(\omega_m^{(0)},0)\omega_m^{(1)}+\partial_{\alpha}\partial_{\omega}\chi^{''}_{aa}(\omega_m^{(0)},0)\}\nonumber\\
&&+\alpha^2\{
\partial^2_{\omega} \chi^{''}_{aa}(\omega_m^{(0)},0)\omega_m^{(2)}+\frac12\partial^3_{\omega} \chi^{''}_{aa}(\omega_m^{(0)},0)(\omega_m^{(1)})^2
\nonumber \\
&&
+ \partial_{\alpha}\partial^2_{\omega} \chi^{''}_{aa}(\omega_m^{(0)},0)\omega_m^{(1)}
+\frac12\partial_{\alpha}^2\partial_{\omega} \chi^{''}_{aa}(\omega_m^0,0)\}+\cdots.
\end{eqnarray}
Solving eq.~(\ref{eq:expansion}) order by order with respect to $\alpha$,
we obtain
\begin{eqnarray}
\omega_m^{(1)}&=&-\frac{\partial_{\alpha}\partial_{\omega}\chi^{''}_{aa}(\omega_m^{(0)},0)}{\partial^2_{\omega} \chi^{''}_{aa}(\omega_m^{(0)},0)}
\label{eq:om_m^1}
\\
\omega_m^{(2)}&=&
-\frac{1}{2 \partial^2_{\omega} \chi^{''}_{aa}(\omega_m^{(0)},0)} 
\big(
\partial^3_{\omega} \chi^{''}_{aa}(\omega_m^{(0)},0)
(\omega_m^{(1)})^2 \nonumber \\
&&
+ 2 \partial_{\alpha}\partial^2_{\omega}
\chi^{''}_{aa}(\omega_m^{(0)},0)\omega_m^{(1)}+
\partial_{\alpha}^2\partial_{\omega} \chi^{''}_{aa}(\omega_m^0,0)
\big).
\nonumber \\
\label{eq:om_m^2}
\end{eqnarray}
To evaluate these expressions,
it is useful to derive the
{\em exact} identities such as
\begin{equation}
\chi_{+-}(\omega) =
- \frac{2\langle S^z \rangle}{N(\omega-H+i\epsilon)}
+ \frac{\langle [ \calA,S^- ]\rangle / N
     + \chi_{\calA \calA^{\dagger}}(\omega)}{
     (\omega - H + i \epsilon)^2},
\label{eq:chi+-id}
\end{equation}
where $N$ is the total number of spins in the system,
$\epsilon$ is the
convergence factor mentioned above,
$\chi_{+-}$ is the complex dynamical susceptibility 
of $S^+$ and $S^-$, and $\chi_{\calA \calA^{\dagger}}$
is that of $\calA$ and $\calA^\dagger$ where $\calA \equiv [\calH', S^+]$.
The identity~(\ref{eq:chi+-id})
follows from the Heisenberg equation of motion
\begin{equation}
\frac{d S^+}{dt} = - iH S^+ + i \calA, 
\end{equation}
and its Hermitian conjugate
(see also Appendix of ref.~\cite{OshikawaAffleck}).
As discussed above, we regard the convergence factor $\epsilon$
as being finite until the expansion ~(\ref{eq:expomega_m}) of the shift
is obtained.

After straightforward but tedious calculations, we obtain
the shift of the main paramagnetic resonance up to the second order.
The first order result reads
\begin{equation}
\alpha\omega_m^{(1)} =
-\frac{\langle[\calA,S^-]\rangle_0}{2\langle S^z\rangle_0},
\label{eq:shift1st}
\end{equation}
where we have introduced the notation $\langle \rangle_n$ as
the $n$-th order contribution (in $\alpha$) to the expectation value
defined with respect to the full Hamiltonian~(\ref{eq:fullHam}).
This result is independent of the polarization.
Although it is found to be identical to the well-known formula by
Kanamori and Tachiki and others~\cite{KanamoriTachiki, Nagata},
our derivation is more systematic than the previous ones.

The second order result, on the other hand, is more
involved and interesting as shown below.
Unlike in the first order term~(\ref{eq:shift1st}),
the shift in second order generally has a polarization dependence.
Thus here we restore the polarization index $a$
in the shift $\delta \omega_{m,a}$, which is given as
\begin{eqnarray}
\lefteqn{
\alpha\omega_{m,a}^{(1)}+
\alpha^2\omega_{m,a}^{(2)}=} \nonumber\\ 
&&
-\frac{\langle[\calA,S^-]\rangle_0+\langle[\calA,S^-]\rangle_1}{2\langle S^z\rangle_0} \nonumber \\
&& +
\frac{\langle[\calA,S^-]\rangle_0}{2\langle S^z\rangle_0}
\left[\frac{\langle S^z\rangle_1}
{\langle S^z\rangle_0} \pm
\frac{\langle[\calA^\dagger,S^-]\rangle_0-\langle[\calA,S^+]\rangle_0}
{4H\langle S^z\rangle_0}\right]\nonumber\\
&&-\frac{N}{8\langle S^z \rangle_0}\lim_{\epsilon\to0}
\biggl[4\chi'_{\calA\calA^\dagger}+
2\epsilon\partial_{\omega}\chi^{''}_{\calA\calA^\dagger}
\nonumber \\
&& \pm \frac1H\left(\epsilon\left(\chi^{''}_{\calA\calA}+
\chi^{''}_{\calA^\dagger \calA^\dagger}\right)-\epsilon^2\left(
\partial_{\omega}\chi'_{\calA\calA}+\partial_{\omega}
\chi'_{\calA^\dagger \calA^\dagger}\right)\right)
\nonumber\\
&&
+ \mbox{``}O(\epsilon^3)\mbox{''} \biggr] ,
\label{eq:shift2nd}
\end{eqnarray}
where $\chi'$  is the real part of the complex dynamical susceptibility,
$\chi'$ and $\chi''$ are evaluated at $\omega=\omega_m^{(0)}$,
and the double sign $\pm$ takes $+$ or $-$ for $a=x$ and $y$,
respectively; $a$ is the polarization direction of the oscillating
magnetic field.
The term $\mbox{``}O(\epsilon^3)\mbox{''}$ refers to the contributions
with prefactors explicitly containing $\epsilon^3$ or higher powers
of $\epsilon$.

Because of the limit $\epsilon \rightarrow +0$,
the terms explicitly multiplied by powers of $\epsilon$, including
the $\mbox{``}O(\epsilon^3)\mbox{''}$ terms, appear to vanish.
However, the correlation functions
in eq.~(\ref{eq:shift2nd}) can have singularities as powers of
$1/\epsilon$, leaving a finite contribution together with the
explicit powers of $\epsilon$.
Thus we cannot ignore such terms a priori, and we 
show later a concrete example with such a nontrivial contribution.
For brevity, in eq.~(\ref{eq:shift2nd})
we have omitted writing out the $\mbox{``}O(\epsilon^3)\mbox{''}$ terms explicitly,
since they actually vanish in all the examples we have studied so far.

Eq.~(\ref{eq:shift2nd}) should be valid up to the second order
in the perturbation parameter $\alpha$,
in any dimension, at any temperature, for any spin, and for any
interaction strength.
However, it must be noted that
the evaluation of the correlation functions appearing
in eq.~(\ref{eq:shift2nd})
is still a nontrivial problem.

\noindent {\em Applications ---}
As a demonstration, here we apply our general second-order result~(\ref{eq:shift2nd})
to the high field limit $\calH_Z \gg \calH_{SU(2)}, \calH'$.
In this case, 
the isotropic interaction $\calH_{SU(2)}$
(if any) may be included in the perturbation $\calH'$.
This is convenient for calculation, because
the unperturbed Hamiltonian $\calH_0=\calH_Z$ is trivially solvable.
As a consequence, eq.~(\ref{eq:shift2nd}) can be evaluated exactly.
Although this is a weak-coupling limit, the problem of the
ESR lineshape is by no means trivial, as is evident in the
results to be shown. 
Here we study a few one-dimensional models, for which we can verify our
results by comparing them to known exact solutions in certain limits.

First we consider the $S=1/2$ XXZ chain in the high field limit.
Here, as discussed above, all the interactions including the isotropic
part are included in the perturbation:
\begin{equation}
\calH'=J\sum_j \mbox{\boldmath $S$}_j \cdot \mbox{\boldmath $S$}_{j+1}-J'\sum_j S^z_jS^z_{j+1},
\label{eq:perurbationXXZ}
\end{equation}
where $J,J' \ll H$ and $z$ is assumed to be identical to the
direction of the applied magnetic field.
For this model, in eq.~(\ref{eq:shift2nd})
the terms involving dynamical susceptibility turn out to vanish,
and contributions from the static expectation values give
the final result:
\begin{eqnarray}
\delta\omega_{m,a}(\alpha) &=&
J'\tanh\left(\frac{\beta H}2\right)
+J'^2\frac{\beta\coth\left(\beta H\right)}{1+\cosh\left(\beta H\right)}
\nonumber \\
&&
-JJ'\frac\beta4{\rm sech}^2\left(\frac{\beta H}2\right)\tanh\left(\frac{\beta H}2\right),
\label{eq:XXZshift}
\end{eqnarray}
where $\beta$ is the inverse temperature.
Because of the U(1) rotational symmetry
about the $z$ axis, there is no polarization dependence in this model.

In the zero temperature limit $\beta \rightarrow \infty$,
eq.~(\ref{eq:XXZshift}) reduces to $J'$.
In this limit, the ESR probes an excitation generated out of the ground state.
Since we consider the strong field case $H \gg J, J'$, the ground state
is simply the fully polarized state and the elementary excitation is
given by a magnon corresponding to single spin flip.
Thus, in the zero temperature limit, the ESR spectrum consists
of a single delta function at $\omega=H+J'$, corresponding
to the magnon energy at the zero wave vector.
Namely, the shift is $J'$, exactly as obtained in our approach.

In the infinite temperature limit $\beta \rightarrow 0$,
eq.~(\ref{eq:XXZshift}) reduces to ${J'}^2/(2H)$.
We emphasize that the ESR lineshape in the infinite temperature limit
$\beta \rightarrow 0$ is a nontrivial problem, although all the
static correlation functions become trivial.
In fact, to our knowledge, no exact result is
available for the general $J/J'$.
Nevertheless, for the special case of $J=J'$, the model reduces to the
XX model, in which 
the infinite temperature limit of the lineshape is
known exactly~\cite{Brandt,Brandt_77}.
(See also ref.~\cite{Maeda:XY} for comparison to the Kubo-Tomita theory
and to a numerical calculation.)
We confirmed that the shift obtained from this exact lineshape
agrees with our result above.
Thus, our result~(\ref{eq:XXZshift})
is verified at both the zero and infinite temperature limits.
We believe that eq.~(\ref{eq:XXZshift}) is exact as the second order
perturbation expansion at any temperature.

Next we turn to the transverse Ising model in the high field limit.
We regard the Ising interaction
\begin{equation}
\calH'=J'\sum_j S_j^xS_{j+1}^x
\end{equation}
as the perturbation of the Zeeman term.
Note that the quantization axis of the Ising interaction is
the $x$-axis and thus the model is the transverse Ising model.
This model lacks U(1) rotational symmetry about the $z$-axis.
Thus it is possible to have a polarization dependence, which
we have indeed obtained. Technically, the evaluation of eq.~(\ref{eq:shift2nd})
for this model is more involved than in
the XXZ model discussed above, as several dynamical
correlation functions such as $\chi''_{\calA \calA}$
give non-vanishing contributions to the shift.
In particular, we find nonvanishing contributions
from the terms 
\begin{equation}
\epsilon\left(
\chi^{''}_{\calA\calA}+\chi^{''}_{\calA^\dagger \calA^\dagger}
\right)
- \epsilon^2\left(
\partial_{\omega}\chi'_{\calA\calA}+\partial_{\omega}
\chi'_{\calA^\dagger \calA^\dagger}
\right)
\label{eq:nontriv}
\end{equation}
in the $\epsilon\to0$
limit, due to the singularity in the dynamical correlations.
Collecting all of the contributions, eq.~(\ref{eq:shift2nd}) for the
transverse Ising model reads
\begin{eqnarray}
\delta\omega_{m,x} &=&
\frac{J'}2\tanh\left(\frac{\beta H}2\right)
\nonumber \\
&& +\frac{J'^2}{16H}\left[
-1+\tanh^2\left(\frac{\beta H}2\right)+\frac{4\beta H}{e^{\beta H}-e^{-\beta H}}
\right] ,\nonumber\\
\label{eq:resultIsingx}
\\
\delta\omega_{m,y}
&=& \frac{J'}2\tanh\left(\frac{\beta H}2\right)
\nonumber \\
&& +\frac{J'^2}{16H}\left[
3-3\tanh^2\left(\frac{\beta H}2\right)+\frac{4\beta H}{e^{\beta H}-e^{-\beta H}\
}
\right],\nonumber\\
\label{eq:resultIsingy}
\end{eqnarray}
for the linear polarization parallel to the $x$ and $y$ axes, respectively. 
The nontrivial polarization dependence is evident in the result.

In the infinite temperature limit $\beta \to 0$,
they are further simplified as
$\delta\omega_{m,x} \to J'^2/(16H)$ and
$\delta\omega_{m,y} \to 5J'^2/(16H)$.
Namely, there is a difference of factor $5$ between the
different polarizations.
They, including the polarization dependence,
are again found to agree with the known exact result~\cite{Capel}
on the transverse Ising model in the infinite temperature limit.
In this case, the polarization dependence, which is necessary for the
agreement with the exact result, comes entirely from
the nontrivial contribution~(\ref{eq:nontriv}).

On the other hand,
our exact second-order results are obtained for the
entire temperature range.
In contrast, the Kubo-Tomita theory and its modification~\cite{Natsume}
assumes a high temperature,
and the previous approach~\cite{McMillan,McMillan_J_60,McMillan_J_61}
based on the projected moment fails to give the polarization dependence.

\noindent {\em Summary --- }
To summarize, we have developed a direct perturbation
theory on the shift of the main paramagnetic resonance in ESR,
and have obtained a general and explicit formula up to the second order.
As a demonstration, it is applied to the $S=1/2$ XXZ chain and
the transverse Ising chain in the strong field regime, and
the shift in the second order
is obtained in closed forms for the entire temperature range.
The agreement with the exact results
available in a few limiting cases is verified.

Our method is also applicable
to systems with higher spins and in higher dimensions.
The evaluation of the formulae in these cases
is possible, at least in the high-field limit.
Furthermore, even for strongly interacting systems,
we could utilize various methods such as field theory
and exact solutions to evaluate eq.~(\ref{eq:shift2nd}).

We also expect that the present approach can be extended to
the problem of the ESR width,
as well as other problems of interest in quantum dynamics.
More details and further generalization will be given
in a separate publication~\cite{inprogress}.

\bigskip

It is a pleasure to thank
K. Asano, S. Miyashita, Y. Saiga and
T. Sasamoto for stimulating discussions.
We are also grateful to I. Affleck
for sending us a copy of ref.~\cite{McMillan}.
This work was partially supported
by a Grant-in-Aid for scientific research
and a 21st Century COE Program at
Tokyo Tech ``Nanometer-Scale Quantum Physics'', both from the
Ministry of Education, Culture, Sports, Science and Technology of Japan.

\end{document}